\documentclass[12pt]{article}
\usepackage{appendix}
\usepackage{anysize}
\usepackage{graphics}
\usepackage{color}
\usepackage{amssymb}
\usepackage{graphicx}
\usepackage{epstopdf}
\usepackage[hang,small,bf]{caption}
\usepackage{hyperref}
\usepackage{amsmath}
\usepackage{latexsym}
\usepackage{setspace}
\usepackage{sectsty}

\def \beq{\begin{equation}}
\def \eeq{\end{equation}}
\def \bea{\begin{align}}
\def \eea{\end{align}}

\def\lsim{\mathrel{\rlap{\lower4pt\hbox{\hskip1pt$\sim$}}
    \raise1pt\hbox{$<$}}}                
\def\gsim{\mathrel{\rlap{\lower4pt\hbox{\hskip1pt$\sim$}}
    \raise1pt\hbox{$>$}}}                

\sectionfont{\large\bf}
\subsectionfont{\normalsize\bf}
\subsubsectionfont{\small\bf}
\title{
\vspace*{-1.3cm}
\begin{flushright}
\normalsize{
ANL-HEP-PR-10-62\\
EFI-10-27\\
FERMILAB-PUB-10-460-T}
\end{flushright}
\vspace{0.5cm}
\Large
\textbf{SUSY-Breaking~Parameters~from~RG~Invariants}\\
\textbf{at~the~LHC}
\author{\textbf{Marcela Carena$^{a,b}$, Patrick Draper$^{b}$,}\\
\textbf{Nausheen R.~Shah$^{a}$, and Carlos E.~M.~Wagner$^{b,c,d}$}\\
[1.5cm]
\normalsize\emph{$^a$~Fermi National Accelerator Laboratory, P.~O.~Box 500, Batavia, IL 60510, USA\footnote{http://theory.fnal.gov}}\\
\normalsize\emph{$^b$~Enrico Fermi Institute and $^c$~Kavli Institute for Cosmological Physics,}\\
\normalsize\emph{University of Chicago, Chicago, IL 60637, USA} \\
\normalsize\emph{$^d$~HEP Division, Argonne National Laboratory, 9700 Cass Ave., Argonne, IL 60439, USA}}}
\begin{document}
\nocite{*}
\setcounter{page}{0}
\maketitle
\thispagestyle{empty}
\begin{abstract}
We study Renormalization Group invariant (RGI) quantities in the Minimal Supersymmetric Standard Model and show that they are a powerful and simple instrument for testing high scale models of supersymmetry (SUSY)-breaking. For illustration, we analyze the frameworks of minimal and general gauge mediated (MGM and GGM) SUSY-breaking, with additional arbitrary soft Higgs mass parameters at the messenger scale. We show that if a gaugino and two first generation sfermion soft masses are determined at the LHC, the RGIs lead to MGM sum rules that yield accurate predictions for the other gaugino and first generation soft masses. RGIs can also be used to reconstruct the fundamental MGM parameters (including the messenger scale), calculate the hypercharge $D$-term, and
find relationships among the third generation and Higgs soft masses. We then study the extent to which measurements of the full first generation spectrum at the LHC may distinguish different SUSY-breaking scenarios. In the case of MGM, although most deviations violate the sum rules by more than estimated experimental errors, we find a 1-parameter family of GGM models that satisfy the constraints and produce the same first generation spectrum. The GGM-MGM degeneracy is lifted by differences in the third generation masses and the messenger scales.

\end{abstract}
\thispagestyle{empty}

\newpage

\setcounter{page}{1}

\onehalfspacing

\section*{Introduction}

Supersymmetry (SUSY) is an extension of the Standard Model (SM) space-time symmetry algebra~\cite{PRPLC.110.1}--\cite{Martin:1997ns} that leads to a tightly constrained set of new particles and interactions and addresses a number of open problems in electroweak-scale physics. These problems include the stabilization of the weak/Planck scale hierarchy, the origin of the negative Higgs mass parameter driving electroweak symmetry breaking, and the existence of dark matter. However, if SUSY is discovered in the laboratory, even the Minimal Supersymmetric Standard Model (MSSM) will introduce a significant collection of new Lagrangian parameters to be measured. Many of these parameters are soft masses that explicitly break SUSY and lift the superpartner spectrum above that of the SM particles. 
It is expected that the fundamental source of SUSY-breaking should be spontaneous rather than explicit, and viable phenomenology is most easily achieved if the breaking takes place in a hidden sector of fields that couple to the MSSM only through higher-dimensional operators. These operators may be generated by integrating out degrees of freedom associated with a characteristic ``messenger scale" $M$. Eventually, with experimental input for several soft masses, it will become interesting to look for patterns that encode the origin of these operators and explain precisely how SUSY-breaking is communicated to the MSSM.

There are several approaches to testing hypotheses about the high scale SUSY-breaking theory and reassembling its parameters from low scale data. One standard method is a top-down fit of high scale parameters to the TeV scale measurements. In the top-down procedure, a Monte Carlo scan is performed over the inputs at $M$, the soft parameters are RG-evolved to the TeV scale, observables are calculated, and a $\chi^2$ statistic is computed for each point in the scan~\cite{Lafaye:2004cn}--\cite{Adam:2010uz}. In another method, the bottom-up approach, one starts with low scale soft parameters and RG-evolves them up until they reach a scale where some structure emerges~\cite{Baer:2003wx}--\cite{Carena:1996km}. However, the $\beta$-functions of all soft sfermion and Higgs parameters are sensitive to both the gaugino masses and the hypercharge $D$-term, $D_Y\equiv \rm{Tr}(Ym^2)$, while the third generation soft sfermion $\beta$-functions contain the soft trilinear parameters. Therefore, all the low-scale soft parameters must be measured before bottom-up reconstruction methods can be used reliably.

A third, complimentary method is provided by 1-loop Renormalization Group invariant~(RGI) quantities in the MSSM~\cite{Carena:2010gr}. Given a model for the generation of the soft parameters at $M$, RGIs facilitate the construction of a wide class of sum rules satisfied by the TeV scale masses. These sum rules can be used either to increase confidence that the model is correct, or to predict unmeasured masses from known masses. RGIs can also be used to reconstruct fundamental parameters at the messenger scale. The RGI reconstruction method is entirely algebraic, and most importantly, it can provide considerable information even if some of the RG-coupled masses in the theory are unknown. This feature is most apparent in models with $\lesssim 10$ degrees of freedom at $M$, and occurs for two reasons: some messenger scale parameters can be reconstructed with RGIs that depend only on a limited set of TeV scale masses, and every sum rule can be traded for an unknown TeV scale parameter. Furthermore, the ability to determine messenger scale parameters with RGIs suggests a useful complementarity to the top-down approach:
every parameter that can be constrained directly with the RGI method can have its range considerably reduced in a Monte Carlo scan.


One powerful constraint on the SUSY-breaking mediation mechanism is already known, and comes from the absence of experimental evidence for large flavor-changing neutral currents. Although not the only method, the simplest approach to achieve agreement with limits from flavor physics is to assume flavor-blindness of the soft parameters at the messenger scale~\cite{Ellis:1981ts}--\cite{Foster:2005wb}.  An attractive way to communicate SUSY-breaking such that the soft sfermion masses are automatically flavor-universal is known as gauge mediation~\cite{Giudice:1998bp}--\cite{Wagner:1998vd}, wherein the hidden sector couples to the MSSM only through SM gauge interactions. As the name suggests, assuming a single messenger scale, general gauge mediation (GGM)~\cite{Meade:2008wd} is the most general formulation of gauge mediation. In this class of models the MSSM soft masses are determined at high scales by current-current correlation functions in the hidden sector.

In Ref.~\cite{Carena:2010gr} we studied the application of RGIs to flavor-blind messenger scale models\footnote{For other studies of RGIs and sum rules in supersymmetry, see Ref.~\cite{Meade:2008wd}--\cite{Balazs:2010ha}. Ref.~\cite{Demir:2004aq} also discusses RGIs in variations of the MSSM by the addition of singlets and extra gauge groups. For simplicity, we restrict our attention to RGIs existing strictly in the MSSM.}. We found that under the well-motivated low scale approximations of minimal flavor violation and degenerate first and second generation soft sfermion masses (as well as the assumption of no new sources of $CP$-violation in the sfermion sector), 14 RGIs could be used to test the flavor-blindness hypothesis and reconstitute all high scale soft parameters as functions of a single undetermined scale. We then applied the method of RGIs to GGM and studied the sensitivity with which certain invariants can detect deviations from a GGM pattern of low scale masses. GGM provides a particularly nice illustration of the method, because under certain conditions, there are exactly enough nonzero invariants so that all parameters controlling the soft masses at the messenger scale and the messenger scale itself can be determined.

Although the application to GGM emphasizes the simplicity of the RGI method, the large number of free parameters in the theory obscures the usefulness of the fact that no single invariant depends on all of the soft masses. It is interesting to consider instead what can be done if only a subset of the soft masses are determined. With this assumption, it becomes necessary to consider more restrictive models of SUSY-breaking with fewer parameters. A convenient example is minimal gauge mediation (MGM), the simplest explicit implementation of gauge mediation. In MGM, a single complete $SU(5)$ representation\footnote{A common generalization of MGM is to increase the number of messengers into $N$ ($\mathbf{5}+\bar{\mathbf{5}}$) representations, which alters the relationship between the gaugino and the sfermion masses; we will comment on this possibility further below, but for our purposes in this work we define MGM to be the case $N=1$.} ($\mathbf{5}+\bar{\mathbf{5}}$) of ``messenger'' particles with characteristic mass scale $M$ couple directly to the SUSY-breaking vacuum expectation value in the hidden sector, while coupling only to the gauge sector of the MSSM. Integrating out the messengers at the scale $M$ produces soft masses of the same form as in GGM, but with specific relationships between all the soft masses~(gauginos as well as the sfermions), controlled by only one mass parameter and the gauge couplings at the messenger scale. Additional contributions to the soft SUSY-breaking masses in the Higgs sector may be required by the solution to the $\mu$-problem, and these contributions can be included with the unknown parameters of the model.

In this study we continue the analysis of the RGIs and focus on their use in the case of less-than-complete information about the low energy soft spectrum. For illustration we work in the context of MGM and consider RGIs that are functions only of the gaugino and the first generation soft masses. If a third generation mass goes unmeasured, it will not be possible to test the flavor-blindness hypothesis completely. However, the vanishing of one particular RGI, $D_{\chi_1}$, in addition to two standard RGI sum rules encoding gaugino mass unification, will still provide a strong hint that a gauge-mediated mechanism is at work.
We will use the remaining RGIs to reconstruct fundamental MGM input parameters and to build two new sum rules satisfied by the gaugino and the first generation soft masses in MGM.

In Section 1 we briefly review the RGIs in GGM and MGM and use them to formulate new sum rules in the latter. In Section 2, we study the MGM sum rules from the first of two directions. Analyzing a minimal set of ``realistic'' experimental measurements,  and after making a simple approximation to absorb the bulk of the 2-loop corrections~\cite{Carena:2010gr}, we find that the gluino mass $M_3$, the left-handed squark mass $m_{\tilde{Q}_1}$, and the right-handed selectron mass $m_{\tilde{e}_1}$ are sufficient to predict the rest of the first generation and gaugino MGM soft spectrum using the sum rules. From these measurements we also extract the messenger scale gauge couplings (and thus the messenger scale itself), non-MGM corrections to the Higgs masses at the messenger scale (encoded in $D_Y(M)$), and the values of the hypercharge, baryon number, and lepton number $D$-terms ($D_{Y_{3H}}(M_c)$, $D_{B_3}(M_c)$, and $D_{L_3}(M_c)$) for the third generation and Higgs at the superpartner scale $M_c$. In Section 3 we consider the complementary scenario in which all of the first generation and gaugino masses have been measured. The sum rules can then be checked directly, and the question arises: how well can an MGM mechanism of SUSY-breaking be distinguished from more general gauge-mediated mechanisms? We concentrate in particular on the new sum rules involving the sfermion mass parameters, which are more complex than the well-known sum rules of gaugino mass unification.  A 1-parameter GGM family of deviations from MGM is found that yields the same first generation and gaugino spectrum, and thus does not violate the sum rules. We discuss methods to distinguish these GGM models from MGM. We find formulations of the sum rules that efficiently test  other deviations from MGM
and follow the analytical analysis with a detailed numerical investigation of the constraints. Discussion and conclusions are given in Section 4.

\section{RGIs in General and Minimal Gauge Mediation}

The GGM framework introduces 6 parameters $A_r$ and $B_r$ controlling the soft masses, as well as the messenger scale $M$ at which the sector transmitting SUSY-breaking to the MSSM can be integrated out, for a total of 7 degrees of freedom in the observed soft spectrum at low scales. Additional model parameters include the bilinear Higgs mass term (or, alternatively, the corresponding value of $\tan (\beta)$, the ratio of the Higgs vaccum expectation values) and the soft trilinear couplings, but they do not appear in the RGIs and we will not need to consider them further in our analysis. Explicitly, the gaugino masses are proportional to three constants $B_r$,\footnote{Relative to the definitions of Ref.~\cite{Meade:2008wd} and our previous work~\cite{Carena:2010gr}, for convenience we absorb a factor of the messenger scale $M$ into the definitions of the $B_r$.}
\begin{equation}\label{softMm}
M_r=g_r^2B_r\;,
\end{equation}
and the soft sfermion and Higgs masses are proportional to three additional parameters $A_r$,
\begin{align}
\label{softfm}
m_{\tilde{f}}^2=\sum_{r=1}^3g_r^4C_r(f)A_r\;,
\end{align}
where $r$ runs over the three gauge groups and the $C_r$ are quadratic Casimirs for the sfermion representations.

\begin{table}[!ptbh]
\begin{center}
\caption{1-Loop RG Invariants in the MSSM }
\footnotesize
\begin{tabular}{@{}| c | c | c | c |@{}}
\hline
\hline
&&&\\
\;\;RGI\;\; & Definition in Terms of Soft Masses & MGM($M$) & GGM($M$)   \\ [0.8ex]
 & & &\\
\hline
\hline
&&&\\
$D_{B_{13}}$&$2(m_{\tilde{Q}_1}^2-m_{\tilde{Q}_3}^2)-m_{\tilde{u}_1}^2+m_{\tilde{u}_3}^2-m_{\tilde{d}_1}^2+m_{\tilde{d}_3}^2$&0&0\\ [3ex]
\hline
&&&\\
$D_{L_{13}}$&$2(m_{\tilde{L}_1}^2-m_{\tilde{L}_3}^2)-m_{\tilde{e}_1}^2+m_{\tilde{e}_3}^2$&0&0\\ [3ex]
\hline
&&&\\
$D_{\chi_1}$&$3(3m_{\tilde{d}_1}^2-2(m_{\tilde{Q}_1}^2-m_{\tilde{L}_1}^2)-m_{\tilde{u}_1}^2)-m_{\tilde{e}_1}^2$&0&0\\ [3ex]
\hline&&&\\
$D_{Y_{13H}}$&$\begin{array}{c} m^2_{\tilde{Q}_1}-2m^2_{\tilde{u}_1}+m^2_{\tilde{d}_1}-m^2_{\tilde{L}_1}+m^2_{\tilde{e}_1}\\-\frac{10}{13}\left(m^2_{\tilde{Q}_3}-2m^2_{\tilde{u}_3}+m^2_{\tilde{d}_3}-m^2_{\tilde{L}_3}+m^2_{\tilde{e}_3}+m^2_{H_u}-m^2_{H_d}\right)\end{array}$&$-\frac{10}{13}(\delta_u-\delta_d)$&$-\frac{10}{13}(\delta_u-\delta_d)$\\ [6ex]
\hline
&&&\\
$D_{Z}$&$3(m_{\tilde{d}_3}^2-m_{\tilde{d}_1}^2)+2(m_{\tilde{L}_3}^2-m_{H_d}^2)$&$-2\delta_d$&$-2\delta_d$\\ [3ex]
\hline
&&&\\
$I_{Y\alpha}$&$\left(m^2_{H_u}-m^2_{H_d}+\sum_{gen}(m^2_{\tilde{Q}}-2m^2_{\tilde{u}}+m^2_{\tilde{d}}-m^2_{\tilde{L}}+m^2_{\tilde{e}})\right)/g_1^2$&$\left(\delta_u-\delta_d\right)/g_1^2$&$\left(\delta_u-\delta_d\right)/g_1^2$\\ [3ex]
\hline
&&&\\
$I_{B_r}$&$M_r/g_r^2$&$B$&$B_r$\\ [3ex]
\hline
&&&\\
$I_{M_1}$&$M_1^2-\frac{33}{8}(m_{\tilde{d}_1}^2-m_{\tilde{u}_1}^2-m_{\tilde{e}_1}^2)$&$\frac{38}{5}g_1^4B^2$&$g_1^4\left(B_1^2+\frac{33}{10}A_1\right)$\\ [3ex]
\hline
&&&\\
$I_{M_2}$&$M_2^2+\frac{1}{24}\left(9(m_{\tilde{d}_1}^2-m_{\tilde{u}_1}^2)+16m_{\tilde{L}_1}^2-m_{\tilde{e}_1}^2\right)$&$2g_2^4B^2$&$g_2^4\left(B_2^2+\frac{1}{2}A_2\right)$\\ [3ex]
\hline
&&&\\
$I_{M_3}$&$M_3^2-\frac{3}{16}(5m_{\tilde{d}_1}^2+m_{\tilde{u}_1}^2-m_{\tilde{e}_1}^2)$&$-2g_3^4B^2$&$g_3^4\left(B_3^2-\frac{3}{2}A_3\right)$\\ [3ex]
\hline
&&&\\
$I_{g_2}$&$ 1/g_1^2-33/(5g_2^2)$&$\approx -10.9$&$\approx -10.9$\\ [3ex]
\hline
&&&\\
$I_{g_3}$&$ 1/g_1^2+11/(5g_3^2)$&$\approx 6.2$&$\approx 6.2$\\ [3ex]
\hline
\end{tabular}
\label{table.Inv}
\end{center}
\end{table}

In order to achieve a realistic value for the Higgsino mass parameter $\mu$, GGM may need to be modified by the inclusion of the parameters $\delta_u$ and $\delta_d$, which represent additional contributions to the soft supersymmetry breaking parameters of the Higgs bosons
beyond those given in Eq.~(\ref{softfm}):
\begin{align}
m_{H_u}^2=m_{\tilde{L}_1}^2+\delta_u,\nonumber\\
m_{H_d}^2=m_{\tilde{L}_1}^2+\delta_d.
\end{align}

The 14 1-loop invariants discussed in Ref.~\cite{Carena:2010gr} and their definitions in terms of the gauge couplings and the 15 soft masses at any scale above the heaviest sparticle mass are given in the first two columns of Table~\ref{table.Inv}. The third and fourth column lists their values in terms of MGM and GGM fundamental parameters at the messenger scale. Small deviations will occur at low scales due to effects of higher order corrections to the $\beta$-functions, and we will discuss them further below.

In GGM, with or without the $\delta_u$ and $\delta_d$ modification, the invariants $D_{B_{13}}$, $D_{L_{13}}$, and $D_{\chi_1}$ are zero at $M$. The vanishing of $D_{B_{13}}$ and $D_{L_{13}}$ provides a stringent test of the flavor-blindness hypothesis, while the vanishing of $D_{\chi_1}$ strongly constrains the parameter space consistent with GGM. If $\delta_u\neq\delta_d$, there are precisely 11 nonzero RGIs, and so all 6 $A_r$ and $B_r$ parameters of GGM, as well as $\delta_u$, $\delta_d$, and the 3 gauge couplings at the messenger scale, can be determined from simple algebraic combinations of the invariants. Explicit formulae are given in Ref.~\cite{Carena:2010gr}. This method relies on the ratio $D_{Y_{13H}}/I_{Y_\alpha}$ in order to extract the hypercharge gauge coupling at the messenger scale, which can then be converted into the other high scale gauge couplings with the invariants $I_{g_2}$ and $I_{g_3}$. Using the analytic expression for the integrated gauge coupling 1-loop RGE,
the messenger scale is determined. On the other hand, if $\delta_u=\delta_d$, then there is one less free parameter, but there are two more constraints on the GGM parameter space given by the vanishing of $D_{Y_{13H}}$ and $I_{Y_\alpha}$. Only 9 non-vanishing RGIs are then available for the determination of 10 unknown high energy parameters. Therefore, from the RGIs one can obtain predictions for 9 of the high energy parameters in terms of a single undetermined one, which can be taken to be one of the gauge couplings at the messenger scale, or equivalently the messenger scale itself.

\subsection{First generation masses and RGIs}

We note from the second column of Table \ref{table.Inv} that of the 14 RGIs, 9 depend only on the gaugino masses, the first generation masses, and the gauge couplings: $D_{\chi_1}$, $I_{M_r}$, $I_{B_r}$, and $I_{g_i}$ ($r=1,2,3$ and $i=1,2$). $D_{\chi_1}$ vanishes in gauge mediated models and thus provides an MGM/GGM sum rule, but cannot be used to determine high scale parameters.

Since $m_{\tilde{Q}_1}$ only appears in $D_{\chi_1}$, the 8 non-vanishing RGIs depend on 10 $M_c$ scale values (4 sfermion masses, 3 gaugino masses and 3 gauge couplings).
Although the Higgs mass parameters, $\delta_{u,d}$, affect the spectrum through their contribution to the hypercharge D-term, these 8 RGIs do not depend explicitly on them, and in GGM they are fixed in terms of the 9 parameters~($A_r$, $B_r$, and $g_r(M)$).  Consequently these RGIs link a total of 19 low and high scale parameters. Measurement of the 10 $M_c$ scale masses would allow the reconstruction of 8 of the messenger scale parameters as a function of a single undetermined one, which can be taken to be $g_3(M)$.

On the other hand, MGM is a 4-dimensional subset of the parameter space of GGM defined by 5 constraints:
\begin{align}
A_1=A_2&=A_3\equiv A,\nonumber\\
B_1=B_2&=B_3\equiv B,\nonumber\\
A&=2B^2.
\label{MGMconst}
\end{align}
The relevant parameters of MGM can therefore be taken to be $g_r(M)$, $B$, $\delta_u$, and $\delta_d$. From the third column of Table~\ref{table.Inv} we see that in MGM, the number of non-vanishing first generation + gaugino RGIs is greater than the number of high scale parameters they depend on.
The 8 relevant RGIs are functions of the same 10 $M_c$ scale values as GGM, but are fixed by only 4 messenger scale parameters ($B$ and the 3 gauge couplings), for a total of 14 parameters. Thus, given 6 measurements (3 gauge couplings and 3 masses) at the scale $M_c$, not only can the $B$ and $g_r(M)$ be reconstructed, but the remaining 4 unmeasured low scale masses can also be predicted (the constraint equation $D_{\chi_1}$=0 allows the determination of $m_{\tilde{Q}_1}$ from the other sfermion masses and thus does not modify this counting).

\begin{figure}[!hbt]
\includegraphics[width=0.95\textwidth,trim=.5in 0in 0.1in 1.9in,clip=true]{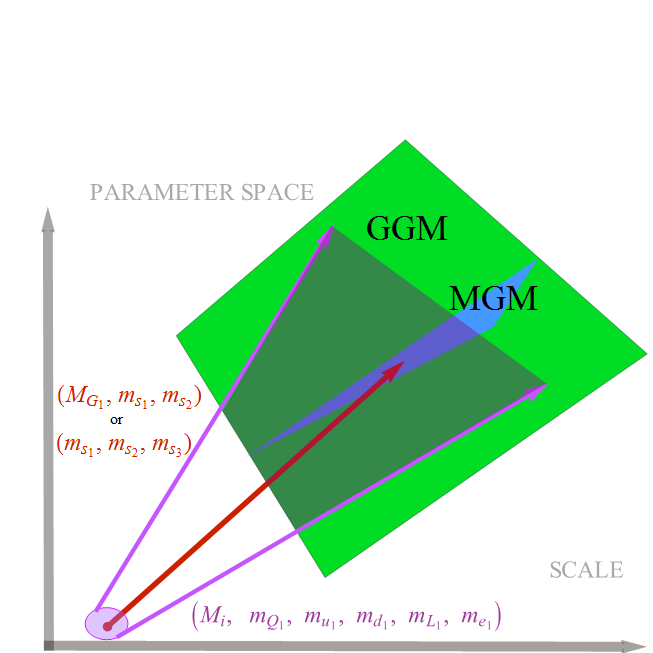}
\caption{The green (light grey) region is GGM parameter space, extending over different $A_r$, $B_r$, $\delta_u$, $\delta_d$, and $M$. The blue (medium grey) region denotes the MGM subspace of GGM, with universal $A_r$ and $B_r$ constrained to satisfy the relationship $A=2B^2$. Using the RGIs and assuming a high scale MGM structure, low scale experimental measurements of only 3 soft masses (small red circle at low scale), including at least 2 scalar masses, can determine consistent $B$ and messenger scale values~(middle red arrow). Low scale measurement of all the gauginos and the first generation masses (purple shaded oval at low scale), on the other hand, leads to the determination of a consistent region of GGM parameter space (shaded region between outer purple arrows).}
\label{pic}
\end{figure}

If the entire first generation spectrum is measured,
the 4 predictions at $M_c$ become sum rules. The equality of the $I_{B_r}$ provide two familiar constraints (and are satisfied more generally in any high scale SUSY-breaking model with gaugino mass unification at the GUT scale). The other two sum rules can be formulated by demanding that the reconstructed gauge couplings at $M$ satisfy the relationships encoded in $I_{g_2}$ and $I_{g_3}$.   These 4 constraints on the low scale soft parameters are related to the 5 constraints given in Eq.~(\ref{MGMconst}). The fifth constraint implied by Eq.~(\ref{MGMconst}) cannot be used to generate a low scale sum rule when considering only the first generation + gaugino RGIs, but instead it allows the extraction of $g_3(M)$ from these RGIs in MGM models. We will discuss the implications of this property in Section 3.
The RGI reconstruction of the high scale parameters of MGM or GGM depends on the parameters that can be measured at the low energy scale, and if the MGM sum rules can be checked.  We depict the different cases described above graphically in Fig.~\ref{pic}.

\section{Predicting an MGM Mass Spectrum}

\subsection{Mass Measurements at the LHC}
In Ref.~\cite{Carena:2010gr}, we analyzed the possibility of distinguishing different SUSY-breaking structures using the RGIs if precise experimental measurements exist for the entire sparticle mass spectrum at the TeV scale. Although this may be possible, it is not the most plausible assumption for the near future. It is more likely that only a subset of all the sparticle and Higgs mass parameters will be determined with good precision at the LHC. Assuming a particular minimal SUSY-breaking scenario, the RGIs can still be used to make predictions for the unmeasured sparticle masses, which can then be tested at a higher luminosity LHC or at future accelerators. To make a detailed program, we try to infer a minimal set of mass measurements that have the most reasonable chance of being performed at the LHC in the coming years.

The LHC will primarily search for supersymmetry by the production of heavy colored particles, which cascade decay into lighter particles. Mass determination of those particles which appear off-shell in the cascade decays will be very difficult, while on-shell particle masses can be determined with relatively good precision. In addition, due to the large backgrounds, the determination of masses in cascade decays containing leptons will be easier to perform compared to those containing only jets plus missing energy (and eventually photons if the messenger scale is low enough). In MGM models, the squarks tend to be heavier than the gluino, and therefore they tend to appear off-shell in the gluino-initiated cascades. Similarly, left-handed sleptons are heavier than the second-lightest neutralino and therefore tend to appear off-shell in cascade decays containing leptons, which will then be dominated by lighter, right-handed sleptons. Cascade decays of gluinos will thus provide information on the gluino mass, the right-handed slepton masses, and the first and second lightest neutralino masses. Further information about the messenger scale may be obtained by the decay of the next-to-lightest superpartner (NLSP) to the gravitino lightest superpartner (LSP), if the messenger scale is low enough ($M \lesssim 10^7$ GeV).

Although produced at a lower rate than gluinos, first and second generation left-handed squarks may be produced at a sufficiently high rate to be measured in the first years of LHC running. These squarks will decay in cascades involving jets, leptons and missing energy. Using the masses obtained in the gluino decays, the left-handed squark masses may be extracted reasonably precisely.

As indicated above, from these cascade decays information on the first and second neutralino masses may be obtained. However, due to the possible mixing of the gauginos with relatively light Higgsinos, these masses will provide only approximate information on the gaugino soft masses $M_1$ and $M_2$. On the contrary, after computing the relevant radiative corrections, the gluino mass will be directly translatable into $M_3$. We shall therefore assume that we have good information on the masses $M_3$, $m_{\tilde{e}_1}$, and $m_{\tilde{Q}_1}$ at the scale of the largest supersymmetric particle mass, which we have denoted $M_c$ as it tends to be the heaviest colored sparticle. The quantum corrections transforming the measured pole masses into running masses at $M_c$ introduce an uncertainty that depends on the unknown spectrum. This problem can be solved by a simple iteration in the calculation of masses, and we shall consider that no significant new uncertainty is induced by the presence of these radiative corrections.

In the last section, we showed that knowledge of three masses and three couplings at the scale $M_c$ are sufficient to determine the MGM parameters as well as the rest of the first generation spectrum. We will then assume that $M_3$, $m_{\tilde{Q}_1}$, and $m_{\tilde{e}_1}$ are measured experimentally with a few percent precision at $M_c$, and that the high-scale SUSY-breaking structure is MGM~\footnote{ In this work, we assume the MGM scenario as an example. It should be noted that similar analysis can be carried out for any choice of SUSY-breaking model.}.

\subsection{MGM Parameters and Sum Rules}

From the the invariants $I_{B_r} = B$ in Table~\ref{table.Inv} and the knowledge of the gauge couplings at $M_c$, we obtain
\begin{equation}
B=\frac{M_3}{g_3^2(M_c)}, \qquad M_1= g_{1}^2(M_c) B,\qquad M_2= g_2^2(M_c) B,
\label{ExtMGMMi}
\end{equation}
which can then be compared with the values obtained from cascade decay measurements. The predictions for $M_{1,2}$ are equivalent to the two sum rules for GUT-scale gaugino mass unification. We stress that in our work we have modified the above predictions with 2-loop corrections to the RGIs, using the simple parametrization given Ref.~\cite{Carena:2010gr}. (Effectively,
in the expressions for $M_1$ and $M_2$ in Eq.~(\ref{ExtMGMMi}), $B$ will be shifted by a term given by the log of an intermediate messenger scale times the difference between the approximate 2-loop $\beta$-functions for the invariants $I_{B_3}$ and $I_{B_r}$.)

From Table~\ref{table.Inv} we see that knowledge of $B$ determines a linear relationship between the invariants $I_{M_r}$ and the gauge couplings $g_r^4(M)$ at the messenger scale. Imposing the relationships encoded in $I_{g_2}$ and $I_{g_3}$ that should be satisfied by the reconstructed messenger scale gauge couplings, we obtain two new sum rules:
\begin{align}
C_1^{\rm MGM} &\equiv\sqrt{\frac{38I_B^2}{5 I_{M_1}}}-\frac{33}{5}\sqrt{\frac{2I_B^2}{I_{M_2}}} - I_{g_2}  \equiv 0,\nonumber\\
C_2^{\rm MGM} &\equiv\sqrt{\frac{38I_B^2}{5 I_{M_1}}}+\frac{11}{5}\sqrt{\frac{-2I_B^2}{I_{M_3}}} - I_{g_3} \equiv 0.
\label{c1c2inv}
\end{align}
Once expressed in terms of low energy mass parameters, Eq.~(\ref{c1c2inv}) together with the constraint
\begin{equation}
D_{\chi_1} = 0
\label{dchi1}
\end{equation}
allow the determination of three sfermion masses in terms of two measured ones.

Additionally, from $I_{M_3}$, $g_3(M)$ satisfies
\begin{equation}
g_3^4(M) = -\frac{I_{M_3}}{2 B^2 }\equiv \frac{1}{2}g_3^4(M_c) \left[\frac{3}{16 M_3^2}\left(5 m_{\tilde{d}_1}^2+m_{\tilde{u}_1}^2-m_{\tilde{e}_1}^2\right)-1\right],
\label{ExtMGMg3}
\end{equation}
while the other two couplings may be determined from the $I_{g_i}$
\begin{align}
g_1^2(M)&= \left( I_{g_3} - \frac{11}{5 g_3^2(M)} \right)^{-1}, \nonumber\\
g_2^2(M)&= \frac{33}{5} \left(I_{g_3} - I_{g_2} - \frac{11}{5 g_3^2(M)} \right)^{-1}.
\label{ExtMGMg1g12}
\end{align}
Thus, once the spectrum is determined from Eqs.~(\ref{ExtMGMMi}), (\ref{c1c2inv}), and (\ref{dchi1}), the messenger scale can be also determined from the measurement of a single gaugino mass and two sfermion masses. Note that instead of using Eq.~(\ref{ExtMGMg3}), one could determine $g_1^4(M)$ or $g_2^4(M)$ through their relationships to the invariants $I_{M_1}$ and $I_{M_2}$. Provided the relationships given in Eq.~(\ref{ExtMGMg1g12}) are fulfilled, any choice would lead to an equivalent result, with errors that will depend on the experimental errors on the associated measured quantities. Finally, as with the gaugino masses, the RGIs given above should be modified to account for the 2-loop corrections. The most important 2-loop effect is in $I_{g_2}$, which can shift by a few percent. We stress that these corrections do not assume the measurement of any additional parameters. We shall also implement them in all subsequent numerical calculations and refer the reader to Ref.~\cite{Carena:2010gr} for details and expressions.

\subsection{Parametric Solutions}

Although the determination of the unknown low scale sfermion masses from Eqs.~(\ref{c1c2inv}) and (\ref{dchi1}) is a generally valid procedure, it is not transparent, since the unknown masses appear in the denominators of square root expressions.
Below we give useful parametric solutions for the predicted masses as functions of $g_r(M)$, which clarifies the dependences on the measured masses and the expected uncertainties in the predictions. The solutions can be easily generalized to cases in which the known masses are different from $M_3$, $m_{\tilde{Q}_1}$ and $m_{\tilde{e}_1}$. For simplicity of presentation, we shall ignore 2-loop corrections.

Once the gaugino masses are determined using Eq.~(\ref{ExtMGMMi}), one can use $D_{\chi_1}=0$ and the $I_{M_{1,2}}$ in Table~\ref{table.Inv} to express the first generation sfermion masses as functions of the measured masses and the gauge couplings at the high scale,

\begin{align}
m_{\tilde{L}_1}^2&=\frac{3}{22}\frac{M_3^2}{ g_3^4(M_c)} \left[g_1^4(M_c)\left(\frac{38}{5} \frac{g_1^4(M)}{g_1^4(M_c)}-1\right)+11g_2^4(M_c)\left(2\frac{g_2^4(M)}{g_2^4(M_c)}-1  \right)\right]-\frac{1}{2}m_{\tilde{e}_1}^2,\label{ExtMGMmL}\\
m_{\tilde{u}_1}^2&= \frac{3}{22} \frac{M_3^2}{ g_3^4(M_c)}\left[\frac{5}{3} g_{1}^4(M_c)\left(\frac{38}{5}\frac{ g_1^4(M)}{g_{1}^4(M_c)}-1\right)-11g_2^4(M_c)\left(2\frac{g_2^4(M)}{g_2^4(M_c)}-1\right)  \right]-\frac{5}{6}m_{\tilde{e}_1}^2+m_{\tilde{Q}_1}^2,\nonumber\\
&&\label{ExtMGMmu}\\
m_{\tilde{d}_1}^2&= -\frac{3}{22}\frac{M_3^2}{ g_3^4(M_c)}\left[\frac{1}{9}g_{1}^4(M_c)\left(\frac{38}{5}\frac{ g_1^4(M)}{g_{1}^4(M_c)}-1\right)+11g_2^4(M_c)\left(2\frac{g_2^4(M)}{g_2^4(M_c)}-1\right)  \right]+\frac{1}{6}m_{\tilde{e}_1}^2+m_{\tilde{Q}_1}^2.\nonumber\\
\label{ExtMGMmd}
\end{align}
Similarly, rewriting Eqs.~(\ref{ExtMGMg1g12}) in terms of the measured gauge couplings and $g_3(M)$,

\begin{eqnarray}
g_1^2(M)&=& g_1^2(M_c) \left[1+\frac{11}{5}\frac{g_1^2(M_c)}{g_3^2(M_c)}\left(1-\frac{g_3^2(M_c)}{g_3^2(M)}\right)\right]^{-1},\nonumber\\
g_2^2(M)&=& g_2^2(M_c)\left[1+\frac{1}{3}\frac{g_2^2(M_c)}{g_3^2(M_c)}\left(1 -\frac{g_3^2(M_c)}{g_3^2(M)}\right)\right]^{-1}.
\label{ExtMGMg1g2g3}
\end{eqnarray}

Substituting in Eq.~(\ref{ExtMGMg3}) the values of $m_{\tilde{u}_1}$ and $m_{\tilde{d}_1}$ from Eq.~(\ref{ExtMGMmu})-(\ref{ExtMGMmd}) and $g_1(M)$ and $g_2(M)$ from Eqs.~(\ref{ExtMGMg1g2g3}), Eq.~(\ref{ExtMGMg3}) becomes a high-degree polynomial equation in $g_3(M)$, with coefficients that are functions of
the measured gaugino and sfermion masses. Ignoring very small terms, it reads
\begin{equation}
g_3^4(M) =\frac{\mathcal{C}}{162} g_3^4(M_c)\left[1- \frac{32}{3~\mathcal{C} }\frac{ g_3^4(M)}{ g_3^4(M_c) }\right] \left[1-\left(1+3\frac{g_3^2(M_c)}{g_2^2(M_c)} \right)\frac{ g_3^2(M)}{g_3^2(M_c)} \right]^2,
\label{Mfg3}
\end{equation}
where
\begin{equation}
\mathcal{C} = \frac{1}{M_3^2} \left( 6 m_{\tilde{Q}_1}^2 - m_{\tilde{e}_1}^2-\frac{5}{33}M_1^2+ 9 M_2^2 - \frac{16}{3} M_3^2  \right).
\end{equation}
Generically, with $M_c\sim\mathcal{O}(1)$TeV, $g_2(M_c)\sim 0.65$ and $g_3(M_c)\sim1.1$. Defining $\chi=g_3^2(M)/g_3^2(M_c)$, Eq.~(\ref{Mfg3}) can be roughly approximated by
\begin{equation}
\frac{162}{\mathcal{C}} \chi^2  \sim \left(1-\frac{32}{3~\mathcal{C}}\chi^2\right)\left(1-9\chi\right)^2.
\label{g3approx}
\end{equation}
We additionally assume that the messenger scale should be in the range $10^5 \lesssim M \lesssim 10^{16}$ GeV and therefore
\begin{equation}
0.25 \lesssim g_3^4(M) \lesssim 1,
\label{g3M}
\end{equation}
or equivalently,
\begin{equation}
0.40 \lesssim \chi \lesssim 0.85.
\label{chiM}
\end{equation}
Note that in MGM, $\mathcal{C}$ is a number of order 10, which becomes somewhat smaller for larger values of the messenger scale. Inspection of Eq.~(\ref{g3approx}) then determines that typically only one $g_3(M)$ solution satisfies Eq.~(\ref{g3M}) and is therefore physically realistic.

It is simple to solve Eq.~(\ref{Mfg3}) numerically, even adding the small terms ignored for simplicity above. It can then be used to calculate the mass spectrum of the first generation by insertion into Eq.~(\ref{ExtMGMmL})-(\ref{ExtMGMmd}).

Furthermore, looking at $D_{Y_{13H}}$ and $I_{Y_\alpha}$, we see that the Higgs and third generation masses appear in the same combination in both RGIs. Therefore, once the quantities in Eqs.~(\ref{ExtMGMMi})-(\ref{ExtMGMg1g12}) have been computed, $D_Y(M)$ and $D_{Y_{3H}}(M_c)$ can be predicted from $D_{Y_{13H}}$ and $I_{Y_\alpha}$:
\begin{eqnarray}
D_{Y}(M)=\delta_u-\delta_d&=& \frac{33}{10}\left(\frac{g_1^2(M_c)}{g_1^2(M)}-1\right)^{-1}(m_{\tilde{Q}_1}^2+m_{\tilde{d}_1}^2-2 m_{\tilde{u}_1}^2-m_{\tilde{L}_1}^2+m_{\tilde{e}_1}^2),\nonumber\\
&=&\frac{33}{10}\left(\frac{g_1^2(M_c)}{g_1^2(M)}-1\right)^{-1} D_{Y_1}(M_c).\label{ExtMGMHiggsDY}
\end{eqnarray}
\begin{eqnarray}
D_{Y_{3H}}(M_c)&=& 2\left(1 +\frac{13}{20} \frac{g_1^2(M_c)}{g_1^2(M)} \right)\left(\frac{g_1^2(M_c)}{g_1^2(M)}-1\right)^{-1}(m_{\tilde{Q}_1}^2+m_{\tilde{d}_1}^2-2 m_{\tilde{u}_1}^2-m_{\tilde{L}_1}^2+m_{\tilde{e}_1}^2),\nonumber\\
&=&\frac{20}{33}\left(1 +\frac{13}{20} \frac{g_1^2(M_c)}{g_1^2(M)} \right)D_Y(M).
\label{ExtMGMHiggs}
\end{eqnarray}
Similarly, since flavor-blindness implies $D_{B_{13}}=D_{B_1}-D_{B_3}=0$ and $D_{L_{13}}=D_{L_1}-D_{L_3}=0$, $D_{B_3}$ and $D_{L_3}$ can be predicted from the assumed measurements:
\begin{align}
D_{B_3}(M_c) & =D_{B_1}(M_c)=2 m_{\tilde{Q}_1}^2-m_{\tilde{u}_1}^2-m_{\tilde{d}_1}^2,\\
D_{L_3}(M_c)& = D_{L_1}(M_c)=2m_{\tilde{L}_1}^2-m_{\tilde{e}_1}^2.
\label{B3L3}
\end{align}
Eqs.~(\ref{ExtMGMHiggs})--(\ref{B3L3}) imply that the determination of 4 soft parameters in the third generation and Higgs sector, in addition to the 3 soft masses of the first generation + gauginos, would be sufficient to fix the soft spectrum entirely.

In Tables \ref{table0} and \ref{table04}, we give two example points in the MGM parameter space where the mass spectrum of the first generation, the messenger scale gauge couplings, $D_Y(M)$, $D_{Y_{3H}}(M_c)$, $D_{B_3}(M_c)$, and $D_{L_3}(M_c)$ are calculated using the equations above. The points correspond to different choices of $A=2B^2$, ($\delta_u-\delta_d$), and the messenger scale. To estimate the uncertainties, we assume that the input sparticle masses \{$M_3$, $m_{\tilde{Q}_1}$, $m_{\tilde{e}_1}$\} have been experimentally measured with central values equal to their MGM values and uncertainties of about 5\%. Although the real errors may be larger than 5\% percent, since we choose a flat uncertainty profile for all the input masses and couplings, the errors in the determined quantities scale roughly linearly with this value, and 5\% provides an easy reference point to establish the re-scaling. For comparison, we also present results for a 1\% uncertainty in the masses, corresponding to future precision measurements.

%
\begin{table}[!htbp]
\begin{center}
\caption{Predicted spectrum of masses and parameters given a minimal set of measurements for the MGM model: $A=2B^2=0.3\mbox{ TeV}^2$, $D_Y(M)=\delta_u-\delta_d=0\mbox{ TeV}^2$, and $M=10^7$ GeV. The scale $M_c=1$ TeV. The Data column gives the model parameters and the associated spectrum obtained by running the soft masses down to the scale $M_c$. The Calculated column gives the predicted mass spectrum and reconstructed model parameters, calculated using Eqs.~(\ref{ExtMGMMi})-(\ref{B3L3}). The final two columns give the estimated experimental uncertainties in the calculated quantites, assuming universal soft mass errors of 1\% and 5\% for the input soft masses.}
\begin{tabular}{||c|c|c|c|c||}
\hline
&&&&\\
& Data & Calculated & $\pm$1\% & $\pm$5\% \\
&&&&\\
\hline
\hline
&&&\multicolumn{2}{|c||} {} \\
$g_1(M_c)$& 0.4693 & &\multicolumn{2}{|c||} {0.0047 (1\%)}  \\
$g_2(M_c)$ & 0.6481 &  & \multicolumn{2}{|c||} {0.0065 (1\%)} \\
$g_3(M_c)$ & 1.0800 &  & \multicolumn{2}{|c||} {0.0108 (1\%)} \\
&&&\multicolumn{2}{|c||} {} \\
\hline
&&&&\\
$M_3$(GeV) & 446.8 & & 4.5 & 22.3\\
$m_{\tilde{Q}_1}$(GeV) &  641.6 & & 6.4 & 32\\
$m_{\tilde{e}_1}$(GeV) &  114.0 & & 1.1 & 5.7\\
&&&&\\
\hline
&&&&\\
$g_1(M)$& 0.5159 & 0.5153 & 0.0093 & 0.0329\\
$g_2(M)$ & 0.6679& 0.6647 & 0.0080 & 0.0131\\
$g_3(M)$ & 0.9144& 0.9093 & 0.0218 & 0.0880\\
&&&&\\
\hline
&&&&\\
$M_1$(GeV)& 84.2 &84.4 & 2.5& 4.9\\
$M_2$(GeV) & 159.4& 158.5 & 4.8& 9.1\\
$m_{\tilde{L}_1}$(GeV)& 227.2 & 221.3 & 10.4& 31.1\\
$m_{\tilde{u}_1}$(GeV) & 608.37& 611.6 & 7.3& 34.8\\
$m_{\tilde{d}_1}$(GeV) & 604.7 & 607.5 & 8.2&38.7\\
&&&&\\
\hline
&&&&\\
$\mbox{Log}_{10}M ~(\mbox{GeV})$ & 7& 6.7&0.6&2.4\\
$A\mbox{ (TeV)}^2$ & 0.3& 0.3 & 0.013 & 0.03\\
$D_Y(M)\mbox{ (TeV)}^2$  & 0& -0.0075 &  0.09&0.31\\
$D_{Y_{3H}}(M_c)\mbox{ (TeV)}^2$& -0.0085 & -0.0130 & 0.08& 0.28\\
$D_{B_{3}}(M_c)\mbox{ (TeV)}^2$& 0.0889 & 0.0904 & 0.0067 & 0.0157\\
$D_{L_{3}}(M_c)\mbox{ (TeV)}^2$& 0.0902 & 0.0933 & 0.0096 & 0.0287\\
&&&&\\
\hline
\hline
\end{tabular}
\label{table0}
\end{center}
\end{table}


\begin{table}[!htbp]
\begin{center}
\caption{Predicted spectrum of masses and parameters given a minimal set of measurements for the MGM model: $A=2B^2=0.8\mbox{ TeV}^2$, $D_Y(M)=\delta_u-\delta_d=0.4\mbox{ TeV}^2$, and $M=10^{12}$ GeV. The scale $M_c=1$ TeV. The Data column gives the model parameters and the associated spectrum obtained by running the soft masses down to the scale $M_c$. The Calculated column gives the predicted mass spectrum and reconstructed model parameters, calculated using Eqs.~(\ref{ExtMGMMi})-(\ref{B3L3}). The final two columns give the estimated experimental uncertainties in the calculated quantites, assuming universal soft mass errors of 1\% and 5\% for the input soft masses.}
\begin{tabular}{||c|c|c|c|c||}
\hline
&&&&\\
& Data & Calculated & $\pm$1\% & $\pm$ 5\% \\
&&&&\\
\hline
\hline
&&&\multicolumn{2}{|c||} {} \\
$g_1(M_c)$& 0.4686   & &\multicolumn{2}{|c||} {0.0047 (1\%)}  \\
$g_2(M_c)$ & 0.6446&  & \multicolumn{2}{|c||} {0.0064 (1\%)} \\
$g_3(M_c)$ & 1.0670 &  & \multicolumn{2}{|c||} {0.0107 (1\%)} \\
&&&\multicolumn{2}{|c||} {} \\
\hline
&&&&\\
$M_3$(GeV) & 707.6  & & 7.1& 35.5\\
$m_{\tilde{Q}_1}$(GeV) &  934.5 & &9.3 & 46.5\\
$m_{\tilde{e}_1}$(GeV) &  228.6 & & 2.3& 11.5 \\
&&&&\\
\hline
&&&&\\
$g_1(M)$ & 0.5981& 0.5881 & 0.0267 & 0.1130\\
$g_2(M)$ & 0.6905& 0.6812 &  0.0112 & 0.0284\\
$g_3(M)$ & 0.7803& 0.7823 & 0.0281 & 0.1511 \\
&&&& \\
\hline
&&&&\\
$M_1$(GeV) & 136.0 & 136.5 & 4.1& 7.8\\
$M_2$(GeV) & 254.0& 254.5 & 7.7 & 14.7\\
$m_{\tilde{L}_1}$(GeV) & 430.5& 409.2 & 32.4& 115.8\\
$m_{\tilde{u}_1}$(GeV) & 869.5& 874.9 & 9.6& 41.9\\
$m_{\tilde{d}_1}$(GeV) & 848.4& 857.0 & 17.5& 76.7\\
&&&&\\
\hline
&&&&\\
$\mbox{Log}_{10}M~(\mbox{GeV})$& 12 & 11.7 & 1.4 & 5.6\\
$A\mbox{ (TeV)}^2$ & 0.8 & 0.78 & 0.035 & 0.086\\
$D_Y(M)\mbox{ (TeV)}^2$ & 0.4& 0.27 & 0.37 & 1.5 \\
$D_{Y_{3H}}(M_c)\mbox{ (TeV)}^2$ & 0.313& 0.22 & 0.31&1.2 \\
$D_{B_{3}}(M_c)\mbox{ (TeV)}^2$& 0.273 & 0.273 & 0.023 & 0.049\\
$D_{L_{3}}(M_c)\mbox{ (TeV)}^2$& 0.316& 0.309 & 0.055 & 0.197\\
&&&&\\
\hline
\hline
\end{tabular}
\label{table04}
\end{center}
\end{table}


The examples in Tables~\ref{table0} and \ref{table04} demonstrate that the method is quite powerful: the propagated errors remain relatively small, and all predicted quantities are within one standard deviation of their true values. The uncertainty in the masses depends on their quantum numbers as well as on the messenger scale. The largest uncertainty is induced by the first terms of Eqs.~(\ref{ExtMGMmL})-(\ref{ExtMGMmd}), associated with the determination of the gauge couplings at the messenger scale. The squark masses, $m_{\tilde{u}_1}$ and $m_{\tilde{d}_1}$, depend only weakly on these terms and predominantly on $m_{\tilde{Q}_1}$. Their relative uncertainty is then small, of the order of a few percent. On the contrary, $m_{\tilde{L}_1}$ depends dominantly on the first term and hence its uncertainty tends to be larger, growing with larger values of the messenger scale. Finally, the messenger scale can be determined. Since the parameters are only mildly logarithmically sensitive to $M$, an accurate determination will demand a precise measurement of the relevant low energy masses. However, we note that the uncertainties in $M$ are much smaller in MGM than what is generically achieved (when reconstructing $M$ is possible) in GGM~\cite{Carena:2010gr}, since fewer parameters are involved and the reconstruction is insensitive to the ($\delta_u-\delta_d$) splitting.

\section{Identifying and Differentiating Minimal within GGM Models}

\subsection{GGM/MGM Models with Degenerate Low Energy Spectra}


If all first generation and gaugino masses are measured at a higher luminosity LHC, the MGM prediction for the spectrum can be tested. However, for each MGM model, there is a corresponding set of non-minimal GGM models which produces the same first generation + gaugino spectrum and thus satisfies the sum rules. These models
form a 1-parameter family which can be parametrized by $g_3(M)$.


To identify the degenerate models, it is convenient to introduce the parameters $x_r$:
\begin{align}
x_r&\equiv A_r/2B_r^2.
\label{xconstrGGM}
\end{align}
Then the GGM input parameters are $B_r$, $x_r$, $\delta_{u,d}$, and the gauge couplings at the messenger scale.
The three parameters $B_r$ can be obtained from the RGIs given by the ratios of the gaugino masses to the gauge couplings squared,
\begin{equation}
B_r = I_{B_r}.
\label{IBrBr}
\end{equation}
Furthermore, the ratios $I_{B_r}^2/I_{M_r}$ constrain the GGM gauge couplings $g_r(M)_{\rm{GGM}}$ and the $x_r$ to satisfy the relationships
\begin{align}
\frac{1}{g_3^{2}(M)_{\rm{GGM}}}&=\left[\frac{-2 I_{B_3}^2}{I_{M_3}}\left(1-\frac{3}{2}(1-x_3)\right)\right]^{1/2},\nonumber\\
\frac{1}{g_2^{2}(M)}_{\rm{GGM}}&=\left[\frac{2  I_{B_2}^2}{I_{M_2}}\left(1-\frac{1}{2}(1-x_2)\right)\right]^{1/2}, \nonumber\\
\frac{1}{g_1^{2}(M)}_{\rm{GGM}}&=\left[\frac{38 \ I_{B_1}^2}{5 \ I_{M_1}}\left(1-\frac{33}{38}(1-x_1)\right)\right]^{1/2},\label{g1mgm}
\end{align}
while Eq.~(\ref{ExtMGMg1g12}) allows the determination of $g_{1,2}(M)_{\rm{GGM}}$ as a function of the $I_{g_r}$ and $g_3(M)_{\rm{GGM}}$. Therefore,
in GGM the
$x_r$ are directly related to the invariants via
\begin{align}
C_1^{\rm{GGM}}  & \equiv
\left[\frac{38 \  I_{B_1}^2}{5 \ I_{M_1}}\left(1-\frac{33}{38}(1-x_1)\right)\right]^{1/2}  -
\frac{33}{5}
\left[\frac{2  I_{B_2}^2}{I_{M_2}}\left(1-\frac{1}{2}(1-x_2)\right)\right]^{1/2}  - I_{g_2}  = 0,
\nonumber\\
C_2^{\rm{GGM}}   & \equiv
\left[\frac{38 \ I_{B_1}^2}{5 I_{M_1}}\left(1-\frac{33}{38}(1-x_1)\right)\right]^{1/2} +
\frac{11}{5} \left[\frac{-2 I_{B_3}^2}{I_{M_3}}\left(1-\frac{3}{2}(1-x_3)\right)\right]^{1/2} - I_{g_3} = 0.
\label{C1C2}
\end{align}
For $x_r = 1$, Eq.~(\ref{C1C2}) becomes the soft mass sum rules of MGM given in Eq.~(\ref{c1c2inv}), but note that (\ref{C1C2}) holds in GGM even if MGM is not a solution.

Given a measured first generation and gaugino spectrum,
the nonzero RGIs are fixed, and thus Eq.~(\ref{C1C2}) defines a curve in the $x_r$ space corresponding to a set of GGM models.  The $A_r$, $B_r$, and $g_r(M)$ parameters of these models are set by Eqs.~(\ref{xconstrGGM})--(\ref{g1mgm}).
Monotonicity implies that the curve can parameterized by (for instance) $g_3(M)_{\rm{GGM}}$. If, in addition, the RGIs satisfy the MGM sum rules in Eq.~(\ref{c1c2inv}), then the curve passes through the point $x_r=1$, and all the models on the curve possess a first generation + gaugino spectrum satisfying the sum rules.

Moreover, these sectors of the spectrum can be made equivalent for all of the models on the curve by changing ($\delta_u-\delta_d$) so that $D_{Y_1}(M_c)$ remains constant. This can be seen as follows. Given values for the 3 $I_{M_r}$ and the constraint $D_{\chi_1}=0$, 4 out of 5 first generation masses can be fixed. If $D_{Y_1}(M_c)$ is also specified, then the fifth mass is fixed. From Eq.~(\ref{ExtMGMHiggsDY}) we see that $D_{Y_1}(M_c)$ is controlled by ($\delta_u-\delta_d$). Therefore, adjusting ($\delta_u-\delta_d$) along the curve can render the models identical in the first generation and gaugino sectors alone\footnote{Note, however, that sufficiently large positive values of ($\delta_u-\delta_d$) may prevent electroweak symmetry breaking at low scales.}.

Consequently, although Eq.~(\ref{c1c2inv}) provides necessary conditions for an MGM spectrum, it is not completely sufficient to rule out more general GGM models. However, Eq.~(\ref{g3M}) limits the physically realizable values of the gauge couplings at the messenger scale, and therefore places bounds on the curve. For completeness, let us mention that for $10^5~{\rm GeV}\lesssim  M \lesssim 10^{16}~{\rm GeV}$, $g_1(M)$ and $g_2(M)$ must lie in the ranges
\begin{align}
0.05\lesssim g_1^4(M) & \lesssim 0.25,\nonumber\\
0.2\lesssim g_2^4(M)  & \lesssim 0.25.
\label{g1g2M}
\end{align}

It is very useful to consider a particular linear combination of the invariants possessing small experimental uncertainties. By themselves the invariants $I_{M_1}$ and $I_{M_2}$ appearing in the constraint functions tend to have large experimental uncertainties: the squark mass terms appear with large coefficients and approximately cancel in gauge mediation, while the experimental uncertainties tend to grow linearly with the squark masses. However, the combination $I_{M_{12}}$, defined as
\begin{align}
I_{M_{12}} & = I_{M_1} + 11 I_{M_2}  \nonumber\\
&   =   M_1^2 + \frac{11}{3}\left(3 M_2^2 + 2 m_{\tilde{L}_1}^2+  m_{\tilde{e}_1}^2 \right)\;,
\end{align}
is manifestly independent of the squark masses, and therefore is likely to be determined more precisely than $I_{M_1}$ or $I_{M_2}$ alone. Since all terms are positive, its fractional error is controlled by the error in the measurement of the weakly interacting sparticle masses. Therefore, in addition to $C_1^{\rm GGM}$ and $C_2^{\rm GGM}$, we will use
\begin{align}
C_5^{\rm GGM}&\equiv\frac{I_{M_{12}}}{I_{B_2}^{2}}+\frac{I_{M_3}}{I_{B_3}^2} \left(1-\frac{3}{2}(1-x_3)\right)^{-1}\nonumber\\
&\times \left\{\frac{95}{11}\frac{I_{B_1}^2}{I_{B_2}^2}\left(1-\frac{33}{38}(1-x_1)\right) \left[11-5I_{g_3}\sqrt{\frac{-I_{M_3}}{ 2\left(1-\frac{3}{2}(1-x_3)\right)I_{B_3}^2}}\;\;\right]^{-2}\right.\nonumber\\
&\qquad\left.+\;1089 \left(1-\frac{1}{2}(1-x_2)\right)\left[11+5\left(I_{g_2}-I_{g_3}\right) \sqrt{\frac{-I_{M_3}}{ 2\left(1-\frac{3}{2}(1-x_3)\right)I_{B_3}^2}}\;\;\right]^{-2}\right\}\nonumber\\
&=0,
\label{C5GGM}
\end{align}
where the constraint function $C_5^{\rm GGM}$ is now a function of $I_{M_{12}}$ and $I_{M_3}$.


\begin{figure*}[!tbh]
\begin{minipage}{1.0\linewidth}
\hspace{-1.5cm}
\begin{center}
\begin{tabular}{cc}
\includegraphics[width=0.5\textwidth]{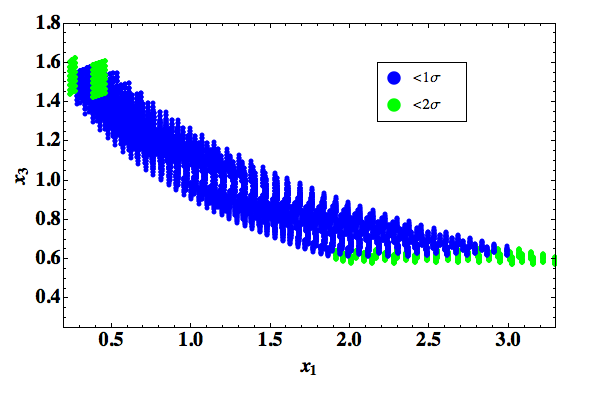} &
\includegraphics[width=0.5\textwidth]{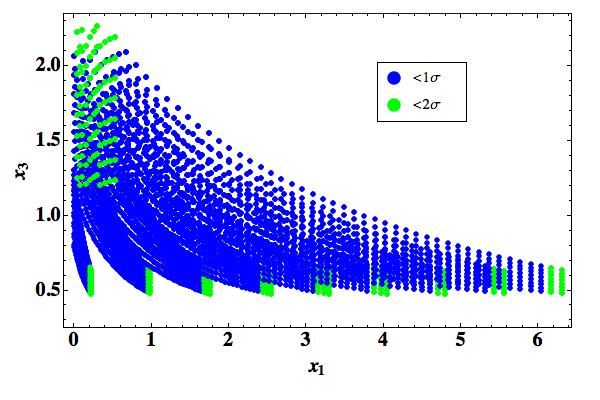}
\end{tabular}
\caption{Constrained $x_1$ vs. $x_3$ GGM parameter space satisfying Eq.~(\ref{C1C2}) and fulfilling the gauge coupling inequalities given in Eqs.~(\ref{g3M}) and~(\ref{g1g2M}) and Eq.~(\ref{C5GGM}) within 1$\sigma$ (Blue/Dark Grey) or  2$\sigma$ (Green/Light Grey) for a sample MGM spectrum with $M=10^{12}$ GeV and $A=2B^2=0.8$. \textit{Left}: 1\% uncertainty; \textit{Right}: 5\% uncertainty.}
\label{GGMxs1}
\end{center}
\end{minipage}
\end{figure*}
\begin{figure*}[!h]
\begin{minipage}{1.0\linewidth}
\hspace{-1.5cm}
\begin{center}
\begin{tabular}{cc}
\includegraphics[width=0.5\textwidth]{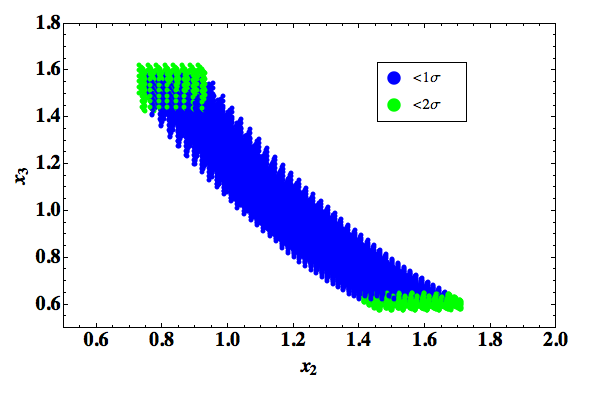} &
\includegraphics[width=0.5\textwidth]{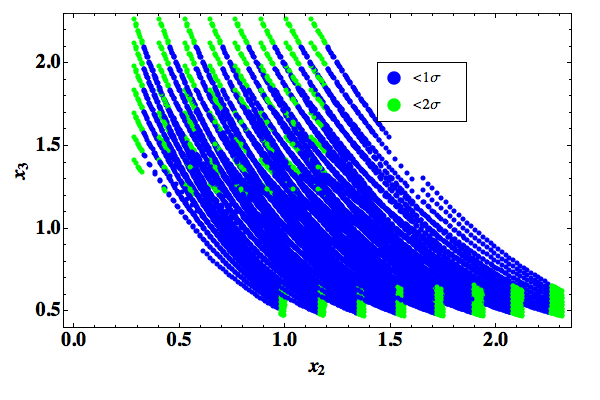}
\end{tabular}
\caption{Constrained $x_2$ vs. $x_3$ GGM parameter space satisfying Eq.~(\ref{C1C2}) and fulfilling the gauge coupling inequalities given in Eqs.~(\ref{g3M}) and~(\ref{g1g2M}) and Eq.~(\ref{C5GGM}) within 1$\sigma$ (Blue/Dark Grey) or  2$\sigma$ (Green/Light Grey) for a sample MGM spectrum with $M=10^{12}$ GeV and $A=2B^2=0.8$. \textit{Left}: 1\% uncertainty; \textit{Right}: 5\% uncertainty.}
\label{GGMxs2}
\end{center}
\end{minipage}
\end{figure*}

Examples of the curve and bounds are given in Figs.~\ref{GGMxs1} and \ref{GGMxs2} for a sample MGM point with $A_r=2B_r^2=0.8\mbox{ TeV}^2$ and $M=10^{12}\mbox{ GeV}^2$. The two plots in each figure reflect an assumption of 1\% and 5\% experimental errors respectively in the measured low scale soft masses. Each point in the $x_r$ space is constrained to satisfy $C_1^{\rm GGM}$, $C_2^{\rm GGM}$, $C_5^{\rm GGM}$ and the gauge coupling inequalities within 1$\sigma$ (Blue/Dark Grey) and 2$\sigma$ (Green/Light Grey). These plots show that while $x_2$ and $x_3$ can in general be well-constrained, the limits on $x_1$ are not as strong. There are two reasons for this behavior. First, a large range of $x_1$ values can be found satisfying $C_1^{\rm GGM}$ and $C_2^{\rm GGM}$ even for values of $x_{2,3}$ relatively close to 1. In other words, Eqs.~(\ref{C1C2}) and~(\ref{C5GGM}) are relatively insensitive to $x_1$. Secondly, from Eq.~(\ref{g1mgm}), large (small) values of $x_1$ correspond to small (large) values of the messenger scale. Therefore, given low (high) scale MGM model, significant low (high) deviations from $x_1=1$ can be tolerated before the upper (lower) bounds of Eq.~(\ref{g1g2M}) are violated.


As explained above, the value of ($\delta_u-\delta_d$) varies along the curve in order to maintain a fixed first generation + gaugino spectrum. Thus the value of the invariant $D_{Y_{13H}}$ is changing, implying that the  third generation + Higgs hypercharge D-term, $D_{Y_{3H}}(M_c)$, is different at each point along the curve. Therefore, the spectrum degeneracy will not hold in the third generation and Higgs sectors, and this fact could be eventually used to select the proper model.

\subsection{GGM Models that can be Distinguished from MGM}

We expect that most deviations from MGM into the more general parameter space of GGM will result in violations of the sum rules. Since the degree of violation should be measured relative to expected uncertainties in the experimental determination of the sum rules, it is important to study them in some detail, looking for alternative formulations that could lead to more stringent constraints.  In this section we will discuss useful reformulations of the constraints $C_1^{\rm MGM}$ and $C_2^{\rm MGM}$ and analyze numerically their ability to rule out deviations from MGM. We will consider simple 1-parameter deviations along different vectors in the $x_r$ space, and then revisit the full class of non-minimal GGM models that satisfy the two constraints.





If Eq.~(\ref{ExtMGMg3}) or Eq.~(\ref{ExtMGMg1g12}) are used to reconstruct an MGM gauge coupling $g_r^2(M')_{\rm MGM}$ at a messenger scale $M'$, but in reality the spectrum is generated by a non-minimal GGM mechanism, then the MGM reconstruction is related to the gauge coupling $g_r^2(M)_{\rm GGM}$ at the real messenger scale $M$ via
\begin{equation}
g_r^2(M')_{\rm MGM}=g_r^2(M)_{\rm GGM}\left[1-c_r(1-x_r)\right]^{1/2},
\label{MMprime}
\end{equation}
where $c_r=\{33/38,~1/2,~3/2\}$.
Correspondingly, if the constraints in Eq.~(\ref{c1c2inv}) are imposed on a non-minimal GGM model, they can be expressed in terms of relations between the messenger scale parameters as
\begin{align}
C_1^{\rm MGM} &\equiv \frac{1}{g_1^2(M)_{\rm GGM}}\left(1-\frac{33}{38}(1-x_1)\right)^{-1/2} -\frac{33}{5}\frac{1}{g_2^2(M)_{\rm GGM}}\left(1-\frac{1}{2}(1-x_2)\right)^{-1/2}-I_{g_2} = 0,   \nonumber\\
C_2^{\rm MGM} &\equiv\frac{1}{g_1^2(M)_{\rm GGM}}\left(1-\frac{33}{38}(1-x_1)\right)^{-1/2} +\frac{11}{5}\frac{1}{g_3^2(M)_{\rm GGM}}\left(1-\frac{3}{2}(1-x_3)\right)^{-1/2}-I_{g_3} = 0.   
\label{c1c2xi}
\end{align}
Since we do not restrict our attention to a fixed low scale spectrum as in the previous section, these equations define a surface in $(M,x_r)$ parameter space containing the line $(M,1,1,1)$ as well as the MGM-degenerate curve discussed in the previous section. For a fixed value of $M$, a new curve within the surface is obtained from Eq.~(\ref{c1c2xi}). We will refer to this curve as the \textit{invariant line}, because it is independent of the way the individual constraint functions are formulated.



In addition to the two ``hard'' constraints of Eq.~(\ref{c1c2xi}), we require that the reconstructed couplings $g_r(M')_{\rm MGM}$ lie in a physically reasonable range. Eqs.~(\ref{g3M}) and (\ref{g1g2M}) become
\begin{align}
0.05\lesssim \frac{5}{38}\frac{I_{M_1}}{I_B^2} \lesssim 0.25\;,\nonumber\\
0.2\lesssim \frac{1}{2}\frac{I_{M_2}}{I_B^2} \lesssim 0.25\;,\nonumber\\
0.25\lesssim -\frac{1}{2}\frac{I_{M_3}}{I_B^2} \lesssim 1.0\;.
\label{gaugeineq}
\end{align}


From the form of Eqs.~(\ref{c1c2xi}) we see that the effectiveness of the constraint functions at detecting deviations from MGM is dependent on the messenger scale through the running gauge couplings.
At low $M$, $g_3(M)_{\rm GGM}$ grows rapidly, $g_2(M)_{\rm GGM}$ decreases slowly, and $g_1(M)_{\rm GGM}$ decreases rapidly, reducing the sensitivities to $x_2$ and $x_3$. On the other hand, the constant coefficients are such that the sensitivity to variations in $x_1$ alone is typically less than to variations in $x_2$ or $x_3$ for all but the lowest messenger scales. Numerically this can be seen by linearizing Eq.~(\ref{c1c2xi}) around $(1,1,1)$ for a sample messenger scale of $10^{5}$ GeV,
\begin{align}
C_1^{\rm MGM} &\approx -1.8 (x_1 -1) + 3.8 (x_2 -1),\nonumber\\
C_2^{\rm MGM} &\approx -1.7 (x_1 -1) - 1.7 (x_3 -1),
\end{align}
and for $10^{16}$ GeV,
\begin{align}
C_1^{\rm MGM} &\approx -0.8 (x_1 -1) + 3.2 (x_2 -1),\nonumber\\
C_2^{\rm MGM} &\approx -0.8 (x_1 -1) - 3.2 (x_3 -1).
\end{align}


Considering the limited reactivity of the constraint functions to $x_1$ and $x_3$ in significant regions of parameter space, it is worthwhile to search for other formulations that have minimal expected experimental uncertainties. For this purpose we write the constraints in a form obtained by constructing the invariants $I_{M_1}/I_B^2$ and $I_{M_2}/I_B^2$ out of the reconstructed $B$ and $g_3^4(M')_{\rm MGM}$. Using Eq.~(\ref{ExtMGMg1g12}) and the relations in Eq.~(\ref{g1mgm}) for $x_r=1$, we arrive at the new constraint functions:
\begin{align}
C_3^{\rm MGM}&\equiv \frac{5I_{M_1}}{38I_B^2}+\frac{50I_{M_3}/I_B^2}{\bigg(22-5I_{g_3}\sqrt{-2I_{M_3}/I_B^2}\bigg)^2},\nonumber\\
C_4^{\rm MGM}&\equiv \frac{I_{M_2}}{2I_B^2}+\frac{2178I_{M_3}/I_B^2}{\bigg(22-5(I_{g_3}-I_{g_2})\sqrt{-2I_{M_3}/I_B^2}\bigg)^2}.
\label{IMiconstr}
 \end{align}
Eq.~(\ref{IMiconstr}) may be rewritten in terms of the $x_r$,
\begin{align}
C_3^{\rm MGM}&=g_1^4(M)_{\rm GGM}\left(1-\frac{33}{38}(1-x_1)\right)-g_3^4(M)_{\rm GGM}\nonumber\\
&\qquad\qquad\;\;\;\times\left(g_3^2(M)_{\rm GGM}I_{g_3}-\frac{11}{5 \left(1-\frac{3}{2}(1-x_3)\right)^{1/2}}\right)^{-2}, \nonumber\\
C_4^{\rm MGM}&=g_2^4(M)_{\rm GGM}\left(1-\frac{1}{2}(1-x_2)\right)-9g_3^4(M)_{\rm GGM}\nonumber\\
&\qquad\qquad\;\;\;\times\left(\frac{5}{11}g_3^2(M)_{\rm GGM}\left(I_{g_3}-I_{g_2}\right)-\frac{1}{\left(1-\frac{3}{2}(1-x_3)\right)^{1/2}}\right)^{-2}.
%
\label{c3c4xi}
\end{align}
Linearizing Eq.~(\ref{c3c4xi}) around $(1,1,1)$, for a messenger scale of $10^{5}$ GeV gives
\begin{align}
C_3^{\rm MGM}&\approx 0.05(x_1 -1) + 0.05 (x_3 -1),\nonumber\\
C_4^{\rm MGM}&\approx 0.09 (x_2 -1) + 0.04 (x_3 -1),
\end{align}
and for $10^{16}$ GeV,
\begin{align}
C_3^{\rm MGM}&\approx 0.2 (x_1 -1) + 0.8 (x_3 -1),\nonumber\\
C_4^{\rm MGM}&\approx 0.1 (x_2 -1) + 0.1 (x_3 -1).
\end{align}

We stress that $C_3^{\rm MGM}$ and $C_4^{\rm MGM}$ are not independent from $C_1^{\rm MGM}$ and $C_2^{\rm MGM}$ on the constraint subsurface defined by $\{C_i=0\}$.
To reduce experimental errors, we can define
\begin{align}
C_5^{\rm MGM}\equiv \frac{19}{5}C_3^{\rm MGM}+11C_4^{\rm MGM}=0,
\label{c5}
\end{align}
which is again a function of $I_{M_{12}}$ and $I_{M_3}$, similarly to $C_5^{\rm GGM}$. For a messenger scale of $10^{5}$ GeV,
\begin{align}
C_5^{\rm MGM}&\approx 0.19(x_1 -1) + 0.99(x_2 -1)+ 0.63(x_3 -1),
\end{align}
and for $10^{16}$ GeV,
\begin{align}
C_5^{\rm MGM}&\approx 0.76 (x_1 -1) + 1.1 (x_2 -1)+ 4.14 (x_3 -1).
\end{align}

Although $C_5^{\rm MGM}$ will prove to be quite a powerful discriminator,
the different constraint functions are sensitive to different deviations from MGM, and so it is still useful to apply all of the $C_i^{\rm MGM}$ as well as the gauge coupling bounds in Eq.~(\ref{gaugeineq}).
For example, $C_3^{\rm MGM}$ involves a square that destroys sign information, so that for low $x_1$ there always exists an $x_3$ that satisfies $C_3^{\rm MGM}=0$. $C_2^{\rm MGM}$, on the other hand, clearly cannot be satisfied for sufficiently low $x_1$, because the first term can become larger than $I_{g_3}$. Secondly,
at large $M$, $C_3^{\rm MGM}$ and $C_5^{\rm MGM}$ lose sensitivity for values of $x_3<1$, which can be anticipated as follows. In this region $g_3(M')_{\rm MGM}$ is less than $g_3(M)_{\rm GGM}$. For high messenger scales, the lower $g_3(M')_{\rm MGM}$ is then translated via $I_{g_3}$ into a reconstructed value of $g_1(M')_{\rm MGM}$ that is increasingly closer to the Landau pole above the GUT scale. Since $C_3^{\rm MGM}$ contains a positive power of the reconstructed $g_1(M')_{\rm MGM}$, the constraint is then violated significantly; however, for constant fractional errors in the low-scale masses, the experimental uncertainties grow even faster (this is just the consequence of the fact that the derivative of a function near a simple pole grows faster than the function itself.) As a result $C_3^{\rm MGM}$ and $C_5^{\rm MGM}$ both become ineffective in this region. On the other hand, precisely because $g_3(M')_{\rm MGM}$ is less than $g_3(M)_{\rm GGM}$, it easily violates Eq.~(\ref{gaugeineq}) when $M$ is large. For these and similar reasons the constraints provide complementarity to each other in different regions of parameter space.


\subsubsection{Numerical Analysis}

We turn now to a numerical study of the constraints and inequalities presented in the previous section. For this purpose we perform scans over GGM parameter space for low and high values of $M$. Since we are interested in the case where ($\delta_u-\delta_d$) is poorly known, and our constraint functions are insensitive to $\delta_u$ and $\delta_d$, we fix them to constant values. We compute the soft spectrum at $M$ and evolve it down numerically to the TeV scale using the full 2-loop RGEs~\cite{Martin:1993zk}. At the TeV scale we assign 5\% uncertainties to each soft mass and add errors in quadrature to obtain final uncertainties on the constraint functions. Although errors in the soft masses may be larger in practice, they are also likely to be highly correlated and may experience cancellations. The simple approximation used here is intended only to provide a qualitative picture of the effectiveness of the RGI method for distinguishing MGM from GGM.

\begin{figure*}[!tbh]
\begin{minipage}{1.0\linewidth}
\begin{center}
\begin{tabular}{cc}
\includegraphics[width=0.5\textwidth]{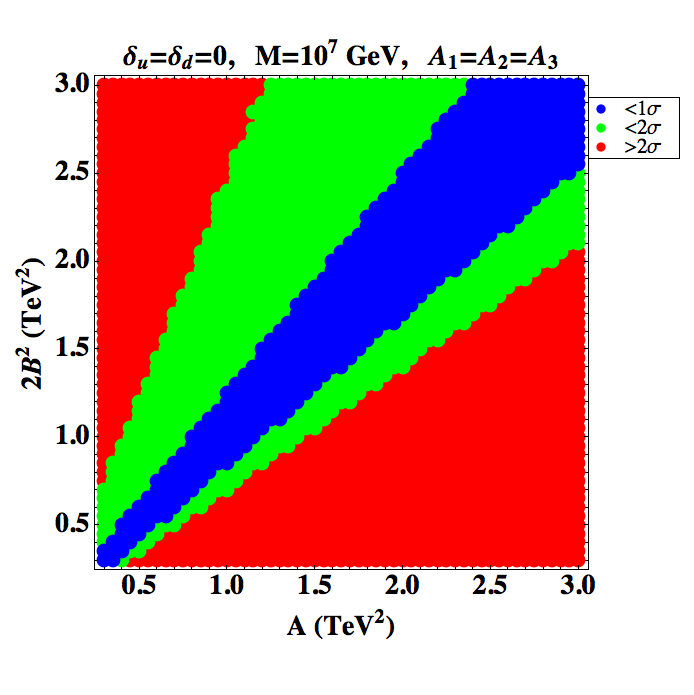} &
\includegraphics[width=0.5\textwidth]{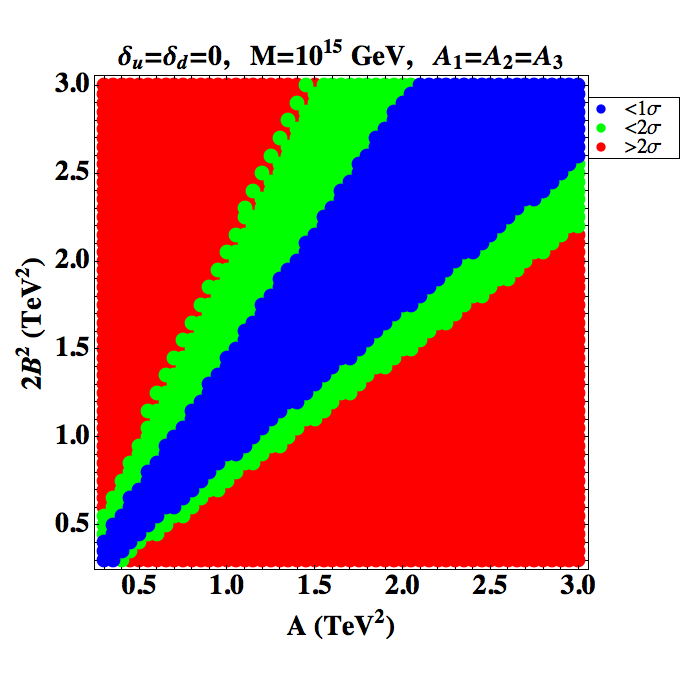}
\end{tabular}
\caption{Ability to rule out MGM in the presence of a deformation in the $A$ direction in GGM parameter space. \textit{Left}: $M=10^7$ GeV; \textit{Right}: $M=10^{15}$ GeV.}
\label{onedevfigA}
\end{center}
\end{minipage}
\end{figure*}

As discussed previously, the simplicity of the $I_{B_r}$ RGIs makes it unlikely that any sizable deviation from $B_1=B_2=B_3\equiv B$ will escape detection once the neutralino spectrum is determined. The $A_r$-dependent constraints are less straightforward, and so as an example we consider deviations from MGM that satisfy universality of the $B_r$ and universality of the $A_r$, but not necessarily $A=2B^2$.


In Fig.~\ref{onedevfigA} we plot the scan points in the GGM subspace, coloring them by the maximum number of standard deviations by which Eqs.~(\ref{c1c2xi}),~(\ref{gaugeineq}), or~(\ref{c5}) are violated. We present results for $M=10^{7}\mbox{ GeV}$ and $M=10^{15}\mbox{ GeV}$, and restrict to a range $0.3<A, 2B^2 < 3.0\mbox{ TeV}^{2}$ for illustration. For these parameter choices, the low scale soft masses range from about 500--2000 GeV for the first generation colored sfermions, 220--900 GeV for the slepton doublet, 110--650 GeV for the slepton singlet, 80--270 GeV for the bino, and 400--1400 GeV for the gluino.


For $x<1$, the sensitivity to displacements from $x=1$ is stronger at large $M$ and is governed by $C_1^{\rm MGM}$, $C_2^{\rm MGM}$, and $g_3(M')_{\rm MGM}$, as discussed in the previous section.  For $x>1$ the dominant constraint comes from $C_5^{\rm MGM}$. Displacements at lower $M$ are also controlled by $C_5^{\rm MGM}$. The slight weakening of the sensitivity towards the upper right in all plots is due to the fact that a constant step size in $A$ or $2B^2$ corresponds to a smaller deviation in $x$ for larger values of $2B^2$.

Note that the space of $x\neq 1$ models includes those that are relevant for $N>1$. If the only deviation from MGM is in the number of messenger multiplets, then the sensitivity reflected in Fig.~\ref{onedevfigA} suggests that a constraint requiring $N$ to be an integer may be reasonably effective. However, we do not investigate this possibility further in this work.

The sensitivity to other types of simple deviations can be understood similarly. Variations purely in $x_1$ for low scales will be the most difficult to detect, as they are weakly felt by $C_1^{\rm MGM}$ and $C_2^{\rm MGM}$ at both scales, while the dependence of $C_5^{\rm MGM}$ on $x_1$ indicates that it is more sensitive at higher scales where $g_1(M)_{\rm GGM}$ is larger. Meanwhile the sensitivity to pure $x_2$ deviations is slightly stronger at low $M$ where the $C_5^{\rm MGM}$ constraint is more powerful.

\begin{figure*}[t]
\hspace{-1.5cm}
\begin{center}
\begin{tabular}{cc}
\includegraphics[width=0.5\textwidth]{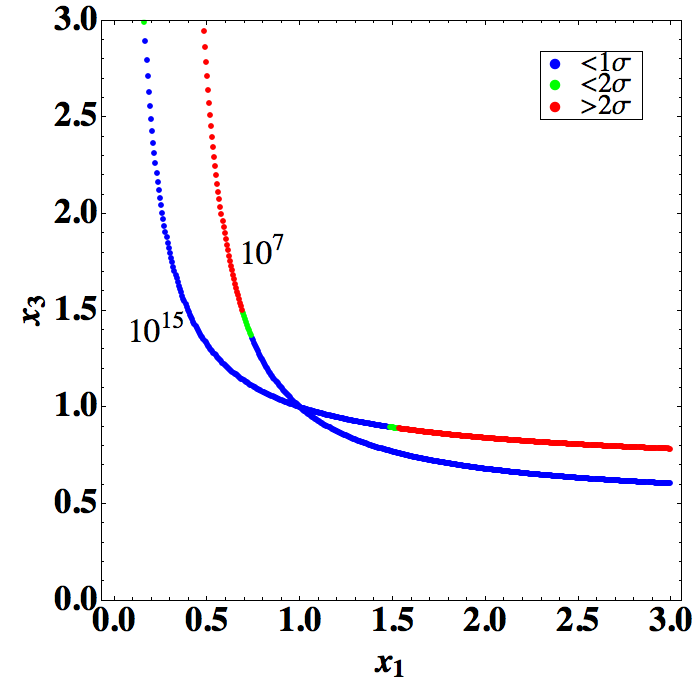} &
\includegraphics[width=0.5\textwidth]{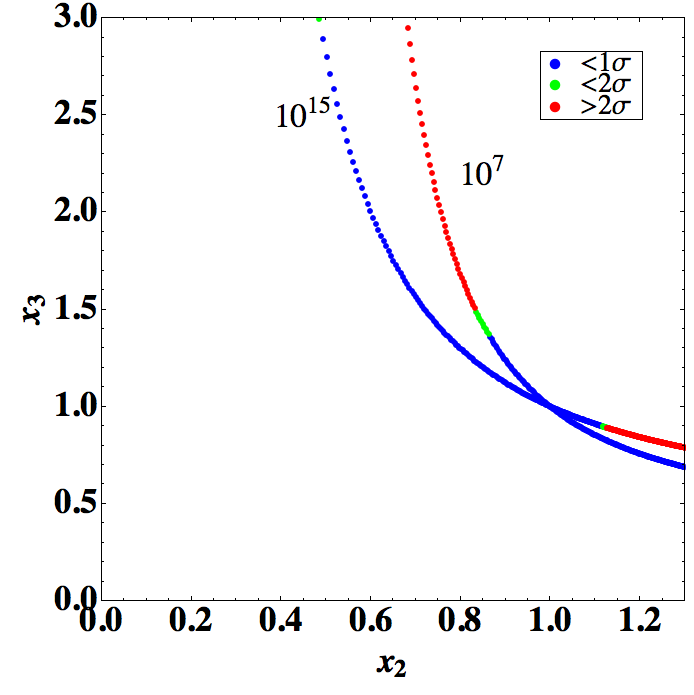}\\
\end{tabular}
\caption{GGM points that survive the constraints may still be distinguished from MGM by demanding that the reconstructed MGM messenger scale is in a physically acceptable range. Here we plot the projections of the unconstrained points onto the $(x_1,x_3)$ plane (\textit{left}) and the $(x_2,x_3)$ plane (\textit{right}), for values of the messenger scale of  $10^7$ GeV and $10^{15}$ GeV. Blue and green (dark and light grey) points remain consistent with the MGM constraints within 2$\sigma$ after the application of the messenger scale inequality; red (medium grey) points are outside the uncertainties. }
\label{invlinefig}
\end{center}
\end{figure*}

While simple deviations are instructive, it is interesting to consider more general displacements from MGM, particularly those that fall directly along the invariant line  As stressed previously, a moderate probe of such cases is offered by enforcing the inequalities in Eq.~(\ref{gaugeineq}).
To estimate the power of the inequalities, in Fig.~\ref{invlinefig} we plot the projection of the invariant line onto the $(x_1,x_3)$ and $(x_2,x_3)$ planes for three values of the messenger scale, coloring points according to whether or not an inequality is violated outside of the error bars. For the estimation of the uncertainties we use a fixed spectrum near $500$ GeV and retain the $5\%$ uncertainties. A more precise calculation does not qualitatively alter the results.
Since values of $x_{1,2}>1$ and $x_3<1$ lead to reconstructed MGM messenger scales that are larger than the true $M$, for low $M$ such deviations are difficult to detect. On the other hand, $x_{1,2}<1$ and $x_3>1$ easily violate the inequalities. At large $M$ the reverse holds since there is not a large margin for the allowed overestimation of the scale.


\section{Conclusions}

In this work we have shown that 1-loop Renormalization Group invariant quantities in the MSSM may be used to study the structure and parameters of SUSY-breaking, even if only a subset of the soft breaking parameters can be determined experimentally.  Working in the specific example of Minimal Gauge Mediation, we found RGI sum rules in the first generation and gaugino sectors that may be used to make predictions for unknown soft masses. We demonstrated that the measurement of one gaugino mass and two first generation sfermion masses at the LHC is sufficient to determine the rest of the first generation and gaugino spectrum in MGM models, three $D$-term relations constraining the third generation and Higgs spectrum, and the high energy input parameters $B$, ($\delta_u - \delta_d$), and the messenger scale.  It is of particular interest  that the relevant RGIs are independent of the Higgs sector soft parameters, the soft trilinear couplings, and all third generation soft masses, which may be more difficult to extract from experimental data.

In the case that the first generation and gaugino masses are known, we showed that the sum rules are sensitive to most deviations into the broader parameter space of General Gauge Mediation, including variations in the sfermion mass parameters that are more complicated to assess than those in the gaugino sector. However, the sum rules cannot completely differentiate MGM from GGM.  A 1-parameter subset of GGM models is consistent with the same low energy first generation and gaugino spectrum as a given MGM model, but is associated with different values of ($\delta_u - \delta_d$) and the messenger scale. Therefore, the GGM models that survive the MGM constraints are limited by the requirement that the reconstructed messenger scale lies within an acceptable window.


It would be interesting to study the breakdown of the degeneracy between MGM and GGM models in the third generation and Higgs spectrum, as well as the determination of ($\delta_u + \delta_d$) in MGM models.  Furthermore, it is of great interest to investigate the complementarity between the RGI method and the top-down approach to SUSY parameter determination. In the event that a parameter can be fixed via RGI relations, it can be restricted in a $\chi^2$ fit, potentially improving the uncertainties from the fit.  Also omitted from our study is a full analysis of other SUSY-breaking scenarios, including MGM with $N>1$ $SU(5)$ representations of messenger particles. We leave such work for the future.

~\\
~\\
{\large{\textbf{Acknowledgements:}}}
\vspace{0.2 cm}

We would like to thank Matt Strassler for useful comments. Fermilab is operated by Fermi Research Alliance, LLC under Contract No. DE-AC02-07CH11359 with the U.S. Department of Energy. Work at ANL is supported in part by the U.S. Department of Energy (DOE), Div.~of HEP, Contract DE-AC02-06CH11357. This work was supported in part by the DOE under Task TeV of contract DE-FGO2-96-ER40956. M.~C., N.~S. and C.~W. would like to thank the Aspen Center for Physics, where part of this work has been done.

\normalsize

%
%
%
%

\end{document}